\documentclass{article}
\usepackage{spconf,amsmath,graphicx}
\usepackage{multirow}
\usepackage{url}
\usepackage{enumitem}
\usepackage{amsmath}
\usepackage{url}

\usepackage{enumitem}
\setlist{nosep, leftmargin=14pt}

\usepackage{mwe} % to get dummy images

\title{{3-D Image-to-Image Fusion in Lightsheet Microscopy by Two-Step Adversarial Network: Contribution to the FuseMyCells Challenge}
\thanks{P\lowercase{ermission from \uppercase{IEEE} must be obtained for all other uses, in any current or future media, including reprinting/republishing this material for advertising or promotional purposes, creating new collective works, for resale or redistribution to servers or lists, or reuse of any copyrighted component of this work in other works.}}}
\name{Marek Wodzinski$^{1,2}$, Henning M\"{u}ller$^{2,3}$}
\address{$^1$Department of Measurement and Electronics, AGH University of Krakow, Krakow, Poland \\
$^2$Institute of Informatics, University of Applied Sciences Western Switzerland, Sierre, Switzerland \\
$^3$Medical Faculty, University of Geneva, Geneva, Switzerland}
\begin{document}
%\ninept
%
\maketitle
\begin{abstract}
Lightsheet microscopy is a powerful 3-D imaging technique that addresses limitations of traditional optical and confocal microscopy but suffers from a low penetration depth and reduced image quality at greater depths. Multiview lightsheet microscopy improves 3-D resolution by combining multiple views but simultaneously increasing the complexity and the photon budget, leading to potential photobleaching and phototoxicity. The FuseMyCells challenge, organized in conjunction with the IEEE ISBI 2025 conference, aims to benchmark deep learning-based solutions for fusing high-quality 3-D volumes from single 3-D views, potentially simplifying procedures and conserving the photon budget.
In this work, we propose a contribution to the FuseMyCells challenge based on a two-step procedure. The first step processes a downsampled version of the image to capture the entire region of interest, while the second step uses a patch-based approach for high-resolution inference, incorporating adversarial loss to enhance visual outcomes. This method addresses challenges related to high data resolution, the necessity of global context, and the preservation of high-frequency details. Experimental results demonstrate the effectiveness of our approach, highlighting its potential to improve 3-D image fusion quality and extend the capabilities of lightsheet microscopy. The average SSIM for the nucleus and membranes is greater than 0.85 and 0.91, respectively.
\end{abstract}
\begin{keywords}
Deep Learning, Image-to-Image, FuseMyCells, Multiview Fusion, Lightsheet Microscopy
\end{keywords}
\section{Introduction}
\label{sec:intro}

Lightsheet microscopy is a 3-D microscopy technique that overcomes issues related to other techniques such as optical or confocal microscopy. Nevertheless, it suffers from a low penetration depth and struggles with accurate imaging of larger samples. The quality, defined as the signal-to-background ratio of the images degrades as the imaging depth increases.

Currently, the problem is solved by multiview lightsheet microscopy that improves the 3-D resolution by combining multiple views of the same sample acquired from different angles. Nevertheless, such a procedure is more complex and increases the photon budget, which can lead to photobleaching and phototoxicity. 

The importance of the task motivated the researchers to organize the FuseMyCells challenge~\cite{FuseMyCells2025}, organized jointly with the IEEE ISBI 2025 conference. The challenge aims to benchmark and compare deep learning-based solutions for automatic fusion of high-quality 3D volumes from single 3D views. The task is important because it may simplify the procedure and save the photon budget. As a result, it may open new possibilities, such as extending the duration of live imaging or adding new fluoresent channels.

Current state-of-the-art methods in 3D image fusion involve various classical and deep learning techniques, such as block matching combined with filtering~\cite{dabov2007image}, content-aware image restoration~\cite{weigert2018content}, self-supervised methods~\cite{krull2019noise2void}, or techniques based on deconvolution~\cite{temerinac2011spatially,toader2022image,prigent2023spitfir}, which have shown promise in improving image quality and resolution.

\begin{figure*}[!ht]
    \centering
    \includegraphics[width=0.70\linewidth]{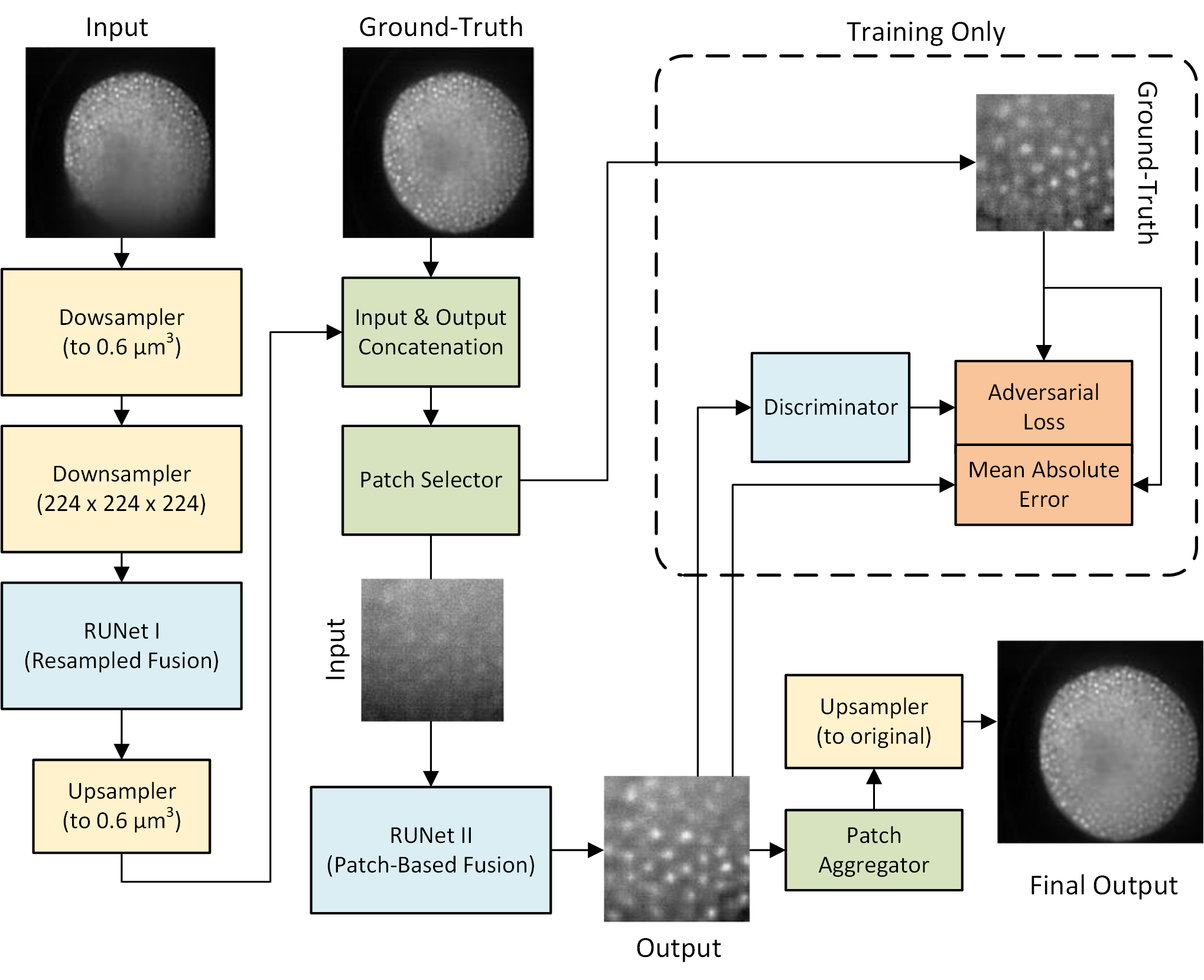}
    \caption{The pipeline of the proposed method.}
    \label{fig:pipeline}
\end{figure*}

Nevertheless, these methods often struggle with the high heterogeneity of the data, the need for high resolution, and the requirement for a global context to accurately fuse images from multiple views.

\textbf{Contribution:} In this work, we present our contribution to the FuseMyCells challenge. We propose a 3D image-to-image fusion method based on a two-step procedure. The first step processes a downsampled version of the image, capturing the entire region of interest. The second step uses a patch-based approach for high-resolution inference, incorporating adversarial loss to improve visual outcomes. The proposed method addresses the challenges related to: (i) high resolution of the data, (ii) the need for the global context, and (iii) the requirement to preserve high-frequency details.

\section{Methods}
\label{sec:methods}

\begin{table*}
\centering
\caption{The quantiative results using the internal validation of the proposed method in terms of MAE, SSIM, nSSIM and nIOU (for the closed test set only). The metrics are calculated using the normalized outputs and ground-truths. The $\pm$ denotes the standard deviation. AL denotes the Adversarial Loss and DW is the weighting based on the absolute difference image.}
\label{tab:evaluation}
\resizebox{\textwidth}{!}{
\begin{tabular}{|l|cccc|cccc|}
\hline
\multirow{2}{*}{\textbf{Method}} & \multicolumn{4}{c|}{\textbf{Nucleus}} & \multicolumn{4}{c|}{\textbf{Membrane}} \\ \cline{2-9} 
 & \textbf{SSIM $\uparrow$} & \textbf{MAE $\downarrow$} & \textbf{nSSIM $\uparrow$} & \textbf{nIOU $\uparrow$} & \textbf{SSIM $\uparrow$} & \textbf{MAE $\downarrow$} & \textbf{nSSIM $\uparrow$} & \textbf{nIOU $\uparrow$} \\ \hline
Identity (Baseline)
& 0.78 $\pm$ 0.20 & 0.22 $\pm$ 0.11 & 0.00 $\pm$ 0.00 & N/A
& 0.70 $\pm$ 0.32 & 0.13 $\pm$ 0.04 & 0.00 $\pm$ 0.00 & N/A \\ \hline
Low Resolution ($224^3$)
& 0.82 $\pm$ 0.12 & 0.15 $\pm$ 0.08 & -0.16 $\pm$ 1.09 & N/A
& 0.76 $\pm$ 0.19 & 0.10 $\pm$ 0.03 & -0.22 $\pm$ 1.89 & N/A \\ \hline
Patch-Based ($224^3$)
& 0.70 $\pm$ 0.21 & 0.13 $\pm$ 0.08 & -0.54 $\pm$ 0.92 & N/A
& 0.69 $\pm$ 0.28 & 0.10 $\pm$ 0.03 & -0.78 $\pm$ 2.39 & N/A \\ \hline
Two Steps
& 0.83 $\pm$ 0.15 & 0.08 $\pm$ 0.08 & -0.09 $\pm$ 1.06 & N/A
& 0.89 $\pm$ 0.11 & 0.06 $\pm$ 0.02 & -0.11 $\pm$ 1.42 & N/A \\ \hline
Two Steps + AL
& 0.83 $\pm$ 0.14 & 0.08 $\pm$ 0.08 & -0.04 $\pm$ 0.96 & N/A
& 0.89 $\pm$ 0.10 & 0.06 $\pm$ 0.03 & -0.02 $\pm$ 1.32 & N/A \\ \hline
Two Steps + AL + DW
& 0.85 $\pm$ 0.11 & 0.07 $\pm$ 0.06 & 0.16 $\pm$ 0.83 & N/A
& 0.91 $\pm$ 0.08 & 0.05 $\pm$ 0.03 & 0.14 $\pm$ 1.51 & N/A \\ \hline
\hline
Two Steps + AL + DW (Closed Test Set)
& N/A & N/A & -0.14 & 0.59
& N/A & N/A & 0.07 & 0.47 \\ \hline
\end{tabular}
}
\end{table*}

\subsection{Overview}

Our method, illustrated in Figure~\ref{fig:pipeline} is s a two-step procedure. The first step processes a downsampled version of the image to fuse the output using the entire region of interest. The second step uses a patch-based approach for high-resolution inference, incorporating the prediction from the previous step and additional adversarial loss.

\subsection{Preprocessing \& Postprocessing \& Augmentation}

The preprocessing includes standardization and resampling to an isotropic spacing of 0.6 $\mu$m x 0.6 $\mu$m x 0.6 $\mu$m. The image is further downsampled to 224x224x224 for the first step of the fusion pipeline.

The postprocessing involves upsampling to the original resolution followed by percentile normalization of the fused image to unify the intensity range. 

During training the initially resampled images are additionally augmented by random cropping and flipping across all axes, to simulate various conditions of quality degradation.

\subsection{Encoder-Decoder Architecture}

Both steps use a traditional 3-D RUNet architecture, chosen based on our previous experience that other encoder-decoder architectures are not beneficial when the size of training data is limited~\cite{wodzinski2024patch}. Moreover, the computational complexity of the task precluded benchmarking several encoder-decoder architectures and performing K-fold cross-validation. Latent diffusion models were also considered but not selected due to their computational demands of the task (maximum allowed inference time equal to 10 minutes per case, including Docker transfer and model loading).

\begin{figure*}[!ht]
    \centering
    \includegraphics[width=0.55\linewidth]{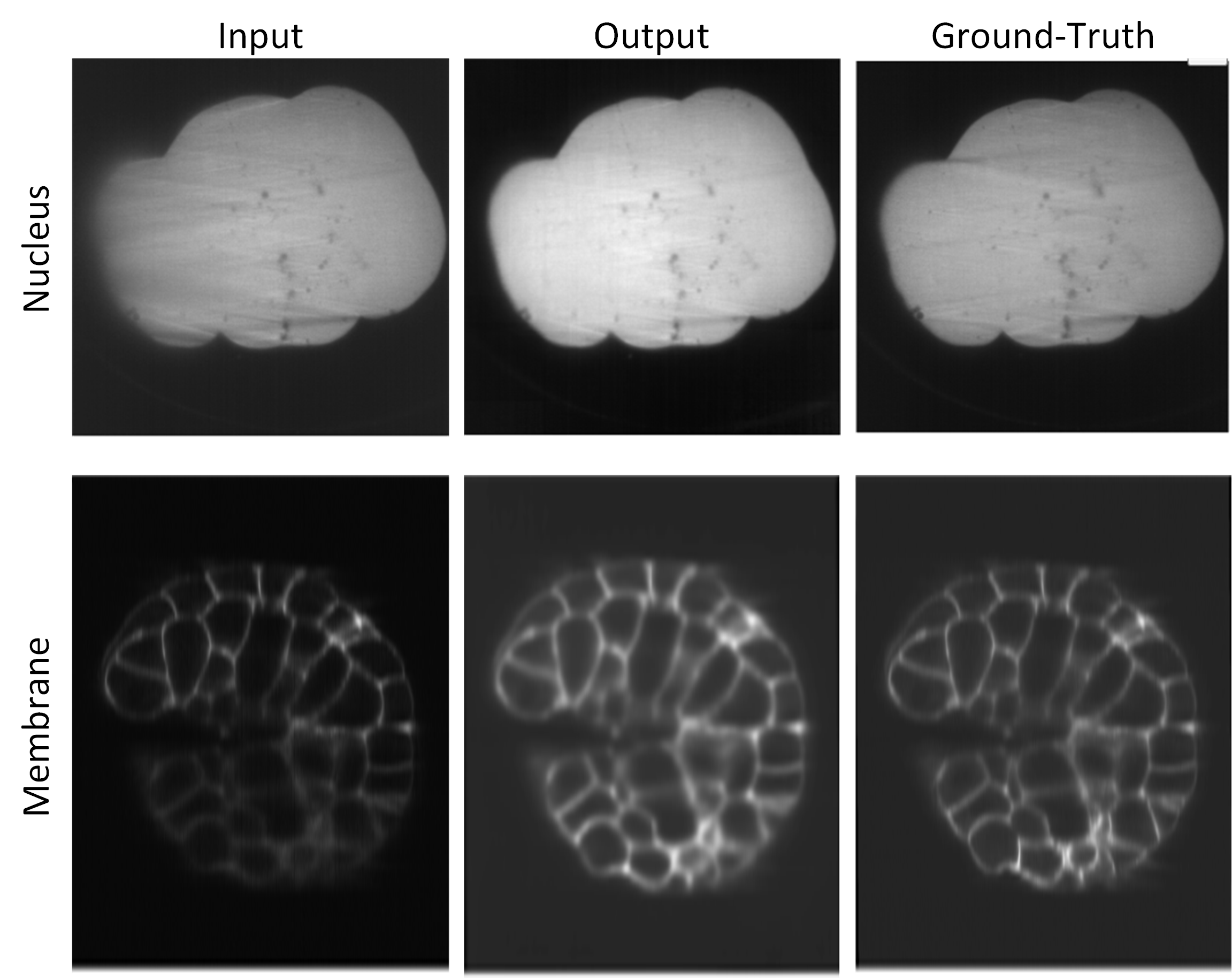}
    \caption{Exemplary outcomes of the proposed method.}
    \label{fig:visualization}
\end{figure*}

\subsection{Training}

The networks were trained using the supervised approach combined with adversarial loss. The objective function was a weighted sum of a mean absolute error (MAE) and an adversarial loss:
\begin{equation}
\begin{split}
L_g(I, Gt) &= \alpha \cdot \text{MAE}(O, Gt) + \beta \cdot \text{AL}(D(O), R) \\
L_d(I, Gt) &= \text{AL}(D(Gt), R) + \text{AL}(D(O), S)
\end{split}
\end{equation}
where $L_g$ is the fusion (generator) loss, $L_d$ is the discriminator loss, $O$ is the fused output, $Gt$ is the ground-truth fused image, $AL$ is the adversarial loss, $\alpha$, $\beta$ controls the relative influence of the optimization objectives, $D$ is the discriminator network, and $R, S$ denote the real and synthetic adversarial expectations, respectively.

Moreover, an additional weighting of the input and output was introduced before the loss calculation to limit hallucinations in parts of the input images with already good views:
\begin{equation}
O = (1 - |I - O|) \cdot I + |I - O| \cdot O
\end{equation},
where $|I - O|$ denotes the [0-1] normalized absolute error between the initial output and the input. During inference, the initial output (before the weightining) was additionally maximum filtered to smooth the prediction.

\subsection{Experimental Setup}

We performed the following ablations: (i) using only low-resolution resampled images, (ii) using only high-resolution patches, (iii) using the two-step approach without adversarial loss, (iv) using the two-step approach with adversarial loss, and (v) using the two-step approach with adversarial loss and difference weighting. Moreover, we defined the baseline as an identity mapping directly returning the input.

All experiments were trained using the AdamW optimizer (two separate optimizers for the ablations with adversarial loss, one for the generator and one for the discriminator) until convergence. 

The experimental setup involved a random 4:1 split for the training and validation subsets. We used PLGRID supercomputer nodes equipped with a 4x NVIDIA GH200 for training and a workstation with 2x A6000 for inference. The implementation was based on the Lightning library~\cite{Falcon_PyTorch_Lightning_2019}. The source code containing the training and inference scripts is openly available~\cite{source_code}.

\section{Results}
\label{sec:results}

We evaluated the experiments using SSIM and MAE separately for the nucleus and membrane, even though one model was used for both. The Table~\ref{tab:evaluation} presents the mean SSIM and MAE for all the experiments. An exemplary visualization of the results is presented in Figure~\ref{fig:visualization}. In addition, we also reported nSSIM and nIOU for the evaluation phase, using the results from the Grand Challenge platform~\cite{FuseMyCells2025}.

The results confirm that the full field of view is crucial for improving the quality of the fusion. Moreover, even though the additional adversarial loss does not improve the results quantitatively, the fine-details are preserved better when compared to the version without the adversarial loss. Finally, the weighting based on the difference between input and output further improves the quality of the fusion.

\section{Discussion}
\label{sec:discussion}

The experimental results demonstrate that the configuration incorporating adversarial loss and difference weighting yielded the most successful outcomes. In particular, the findings underscore the importance of the global context in achieving accurate fusion, as opposed to the traditional patch-based approach, which underperforms even when compared to identity mapping (Table~\ref{tab:evaluation}).

The use of adversarial loss plays an important role in preserving high-frequency details, surpassing the performance of mean absolute error alone. Similarly, difference weighting facilitates the direct propagation of input regions that do not require modification, which is crucial because the absence of this mechanism leads to blurriness in unmodified regions.

A notable limitation of the proposed methodology is its relatively high computational complexity. The training process was conducted using a cluster comprising four NVIDIA GH200 GPUs (each with 98 GB of VRAM), 460 GB of RAM, and efficient disk storage. This level of computational power is essential, as training and executing these methods on more accessible computing resources would be infeasible.

Another limitation is related to the generalizability of the proposed method. The Table~\ref{tab:evaluation} confirms that there is a huge performance gap between the internal validation set and the closed test set. The closed test set partially consists of previously unseen studies, and it may be the main cause of the differences.

It should be noted that we observed limited correlation of the mean SSIM with the mean nSSIM when performing the internal evaluation. The nSSIM was strongly influenced by outliers, and even a single wrong fusion could result in a significant drop in the metric. The Table~\ref{tab:evaluation} confirms that even though the mean SSIM between the identity mapping and the best performing setup increased significantly, the nSSIM still oscillated around 0. Therefore, in the final submission we decided to use model checkpoint that minimizes mean nSSIM, although other checkpoints could provide better results in terms of median nSSIM, overall SSIM, or even potentially nIOU.

In conclusion, this work describes our contribution to the FuseMyCells challenge, which focuses on predicting a fused 3-D image from a single 3-D view. We introduced a two-step approach that leverages the full field of view while preserving fine details. In addition, we improved the method with adversarial loss and difference weighting to improve the fidelity of the reconstruction. Future research could explore nucleus- and membrane-based conditioning to further enhance the fused image quality.

\section{Compliance with ethical standards}
\label{sec:ethics}

This research study was conducted retrospectively using microscopy data made available in open access by FuseMyCells Challenge organizers. Ethical approval was not required, as confirmed by the license attached with the open-access data.

\section{Acknowledgments}
\label{sec:acknowledgments}

The research was partially supported by the program "Excellence Initiative - Research University" for AGH University. We gratefully acknowledge Polish high-performance computing infrastructure PLGrid (HPC Center: ACK Cyfronet AGH) for providing computer facilities and support within computational grant no. PLG/2025/018194.

% ------------------------------------------------------------------------- 
\bibliographystyle{IEEEbib}
\bibliography{strings,refs}

\end{document}